\documentclass[preprint,amsmath,amssymb,aps]{revtex4-1}

\usepackage{graphicx}
\usepackage{dcolumn}
\usepackage{bm}
\usepackage{float}
\usepackage{color}
\usepackage{amssymb}
\usepackage{tikz}
\usetikzlibrary{shapes}
\usepackage{xcolor}
\usepackage{subfig}
\usepackage[percent]{overpic} 
\usepackage{enumitem}

\begin{document}

\title{Reversible bootstrap percolation: Fake news and fact checking}

\author{Mat\'ias A. {Di Muro}}
\affiliation{Instituto de Investigaciones F\'isicas de Mar del Plata
  (IFIMAR)-Departamento de F\'isica, Facultad de Ciencias Exactas y
  Naturales, Universidad Nacional de Mar del Plata-CONICET, Funes
  3350, (7600) Mar del Plata, Argentina.}
\email{mdimuro@mdp.edu.ar}

\author{Sergey V. Buldyrev}
\affiliation{Department of Physics, Yeshiva University, 500 West 185th
  Street, New York, New York 10033, USA}
\affiliation{Politecnico di Milano, Department of Management, Economics and Industrial Engineering,
Via Lambruschini 4, BLD 26, 20156 Milano, Italy}

  
\author{Lidia A. Braunstein}
\affiliation{Instituto de Investigaciones F\'isicas de Mar del Plata
    (IFIMAR)-Departamento de F\'isica, Facultad de Ciencias Exactas y
    Naturales, Universidad Nacional de Mar del Plata-CONICET, Funes
    3350, (7600) Mar del Plata, Argentina.}
  \affiliation{Physics Department, Boston University, Boston,
    Massachusetts 02215, USA}

\begin{abstract}
\noindent 
Bootstrap percolation has been used to describe opinion formation in
society and other social and natural phenomena.  The formal equation
of the bootstrap percolation may have more than one solution,
corresponding to several stable fixed points of the corresponding
iteration process. We construct a reversible bootstrap percolation
process, which converges to these extra solutions displaying a
hysteresis typical of discontinuous phase transitions. This process
provides a reasonable model for fake news spreading and the
effectiveness of fact-checking. We show that sometimes it is not
sufficient to discard all the sources of fake news in order to reverse
the belief of a population that formed under the influence
of these sources.
\end{abstract}
  
\maketitle


The spread of information within a population is an interesting
phenomenon from which we can learn, how prone is a massive group of
people to embrace and propagate fake news or conspiracy theories for instance.
One of the models that can describe this process is bootstrap
percolation, which is a fairly simple threshold model that has been
widely studied in the last years to mimic different spreading
processes on complex systems. In this model a random fraction of nodes
or sites activate or adopt a new idea spontaneously. Then, other nodes subsequently activate if they are connected with at least a
minimal number of active neighbors \cite{Watts}. The initial random
activation triggers a cascading process which stops at when the system
stabilizes.

This model was first introduced to understand the mechanisms of
ferromagnetism on Bethe lattices \cite{Chalupa}, and then in the
following years, it was studied in a variety of graphs
\cite{janson2009percolation,Bax_2010,Bax_2011}. Bootstrap percolation, along with other thresholds
models such as $k$-core percolation, are useful to describe social
processes, in which people tend to change their opinion if they are
influenced by multiple contacts
\cite{centola2010spread,JacksonDiff}. Accordingly, these models can
potentially describe phenomena such as the spreading of gossip or fake
news, viral marketing and opinion formation
\cite{kempe2003maximizing,DomingosMarketing,Shrestha}. Also, people
tend to adopt new technologies or brands when they are in contact with
people that are already using them
\cite{Gleesoncorrelated,Kempe_Inf}.  Bootstrap percolation has also
nonsocial applications, such as the study of fault tolerance in
distributed computing \cite{Kirkpatrick} and cascading failures in
power grids or communication networks.  Furthermore, the spreading of
a disease and the diffusion of awareness \cite{Granell} can be studied using these
kinds of threshold models \cite{Feng}.


In bootstrap percolation, a fraction $f$ of nodes are spontaneously
activated at the initial stage of the process. Such nodes are called
``seeds,''\ while the rest of them are called ``nonseeds.'' A nonseed
node with degree $k$ needs to be supported or influenced by at least
$k^*\leq k$ active neighbors to be activated. The values of $k$ and
$k^\ast$ may be different for different nodes and we will assume that
they are randomly chosen from the degree distribution $P(k)$ and a
threshold distribution $r(j,k)$, respectively, where $r(j,k)$ is the
cumulative distribution function of the threshold which is the probability of
finding a node with $k^\ast \leq j$, given that it has degree $k$. The
activation of the seed nodes leads to a cascade of activation at the
end of which the fraction $S\geq f$ of nodes become active. This
fraction can be regarded as an order parameter of the model, and at
certain $f=f_t$ may undergo a discontinuous transitions similar to
crossing a spinodal associated with the first-order phase transitions in condensed matter, jumping from a small value for $f<f_t$ to a larger value
for $f>f_t$. For example, in the liquid-gas phase transition
there are two spinodals (lines of diverging compressibility) emanating from the critical point\cite{Huangbook}. When approaching spinodals,
the uniform metastable phases (supercooled gas or superheated liquid) reach their stability limits and immediately phase segregate into a mixture of two phases, forming droplets of liquid or bubbles of gas. The
authors in network science call such a transition a hybrid transition \cite{Bax_2010} because approaching this transition from one side the derivative of $dS/df$ diverges and the removal of a single node causes  power-law distributed avalanches, thus they have features of both first- and second-order phase transitions. However the same phenomenology is present near the spinodal: isothermal compressibility $(-\partial V/\partial P)_T/V$ diverges, and the uniform density phase has diverging density fluctuations, before it breaks down and phases segregate. The only difference is that in networks the actual line of the equilibrium first-order phase transition defined by the condition of equal chemical potentials of the two phases is not observable, since the thermodynamic potential is not defined. 

The bootstrap percolation model can be solved exactly in the limit of infinitely large networks randomly connected with a given degree distribution~\cite{Bax_2010,DiMuroBootstrap},
when the probability of short loops is negligible. At the end of
activation cascade for given $f$ the fraction of active nodes can be
written as

\begin{equation}\label{Act}
S=f+(1-f)\Psi(Z),
\end{equation}
where $\Psi(Z)$ is the bootstrap generating function
\begin{equation}
  \Psi(Z)=\sum_k P(k) \sum_{j=0}^{k}r(j,k)\tbinom{k}{j}Z^j(1-Z)^{k-j},
\end{equation}
and probability $Z$ satisfies a self-consistent equation
\begin{equation}\label{self}
Z=f+(1-f)\Phi(Z),
\end{equation}
where 
\begin{equation}
  \Phi(Z)=\sum_k \frac{kP(k)}{\langle k \rangle} \sum_{j=0}^{k-1}r(j,k)\binom{k-1}{j}Z^j(1-Z)^{k-j-1}
  \label{e:Phi}
\end{equation}
is the bootstrap generating function for the excess degree distribution,
$\langle k \rangle$ is the average degree of the network, and $Z$ is
 the probability of reaching via a random link a seed node or a nonseed node supported by at least $k^*$ of its $k-1$ outgoing neighbors.
 The nonlinear equation (\ref{self}) may have more than one solution, which can be obtained by an iteration method
 corresponding to the stages of the activation/deactivation process:
 \begin{equation}\label{iter}
Z_{n+1}=f+(1-f)\Phi(Z_n),
 \end{equation}
 where $Z_n\to Z$ for $n\to \infty$. For a fixed fraction of seed
 nodes $f$, if the initial value of $Z=Z_0$ of Eq.~(\ref{iter}) is small enough,
 e.g., $Z_0\leq f$, then the iterations converge to the smallest stable fixed
 point $Z_I$ corresponding to the direct bootstrap percolation. On the
 other hand, if the initial value of $Z$ is large enough, e.g., $Z_0>1-\epsilon$, where $\epsilon >0$, the
 iterations converge to a different stable fixed point
 $Z_{II}\geq Z_I$ if such a point exists. For many reasonable degree and threshold distributions, function $y=f+(1-f)\Phi(x)$ has inflection points and crosses the straight line $y=x$ several times, producing several stable fixed points of the iterative process (\ref{iter}), when at the crossing point $(1-f)d\Phi(x)/dx<1$.
 In the limit of an infinitely large network, if $Z_0=0$ and the fraction of active nodes $S_0$ coincides with the fraction of seeds $S_0=f$, equation
 \begin{equation}
   S_n=f+(1-f)\Psi(Z_n)
   \label{e:Sn}
 \end{equation}  
 gives the fraction of active nodes after the $n$th stage of the activation cascade such that at each stage each inactive node $i$ counts its already active neighbors and, if this number $k^a_i\geq k^\ast_i$, node $i$ activates. However, it is not clear, how to solve this problem analytically for other initial conditions of the network.    
 It is tempting to suggest that the second
 fixed point corresponds to a reverse process associated with a
 hysteresis phenomenon typical of discontinuous phase transitions in general and network science in particular \cite{Min2014,Majdandzic2014}. One
 potential candidate could be $k$-core percolation
 \cite{dorogovtsev2006k,doro_2019}. In this process, all nodes are
 initially active when a random fraction $1-p$ of them fails
 spontaneously . Each node has a functionality threshold, i.e., a
 minimum number of neighbors $k_c^\ast$ that must be active to keep it in that
 state. Thus, the random failure generates a pruning process that may
 end up affecting a significant amount of nodes in the
 network. Even though bootstrap and $k$-core are quite similar, it has been shown
 that they are not opposite processes but are complementary under the
 right conditions~\cite{DiMuroBootstrap}. In particular, the $k$-core
 process with $p=1-f$ and a complementary threshold distribution is not a reverse process for the bootstrap
 percolation but describes the decrease of the number of inactive nodes in the same bootstrap percolation process. In general, it is impossible to construct a $k$-core process such that it will describe the reverse bootstrap process.

The problem with Eqs.~(\ref{Act}-\ref{iter}) is that the physical meaning
of the probability Z n is not very clear. In Ref. \cite{DiMuroBootstrap} it was
defined as the probability of reaching an already activated
node or a seed by a randomly selected directed link, emanating
from any node, active, inactive, or seed. But this definition
does not link $Z_n$ to any material feature of the network and
does not specify how it can be computed for a given network
configuration. In contrast, the physical meaning of $S$ is clear.
It is the fraction of active nodes in the system and it is easy
to construct a cascading process of consecutive activation of
nodes in which the initial set of active nodes $A_0$ coincides
with the seed nodes and the set of active nodes A n at the
nth stage of the cascade can be readily determined by the set $A_{n-1}$.

 The aim of this paper is to find the physical meaning of $Z_n$ and construct the inverse process for bootstrap percolation that converges to the larger fixed point of Eq.~(\ref{iter}). In addition, in the case when Eq.~(\ref{self}) has a
 unique solution $Z_s$, the process must reach that solution starting
 from an initial condition $Z_0$ such that $Z_0>Z_s$. To achieve this goal, we develop a reversible bootstrap percolation model, which dynamically responds to the changes in the seed configuration, leading to activation of nodes if the fraction of seeds increases and to deactivation of nodes if the number of seeds decreases.
 We assume that during this process some nodes can spontaneously become
 active without being surrounded with at least $k^\ast$ neighbors, thus becoming
 seeds which are self-sufficient, while some nodes can spontaneously lose this property, i.e., become regular nodes without being immediately deactivated. From the social perspective, these nodes change from agents actively influencing their neighbors by spreading news or advertising new products (for example, because they are paid for doing this) to regular members of a community whose beliefs and habits are influenced solely by the beliefs and habits of their contacts in the social network. The reduction of the number of seed nodes and the subsequent deactivation of other nodes can be associated with the process of fact checking.
 
 We must state that our current model may not be adequate for real social
 phenomena, because we deliberately assume that the network topology does not change during this process. In reality, this may not be true. The agents involved in these activities can be simply deleted from the network or inserted into it. We ignore such events, to keep the model analytically tractable.

\begin{figure}[H]
\begin{center}
  \includegraphics[width=0.47\textwidth]{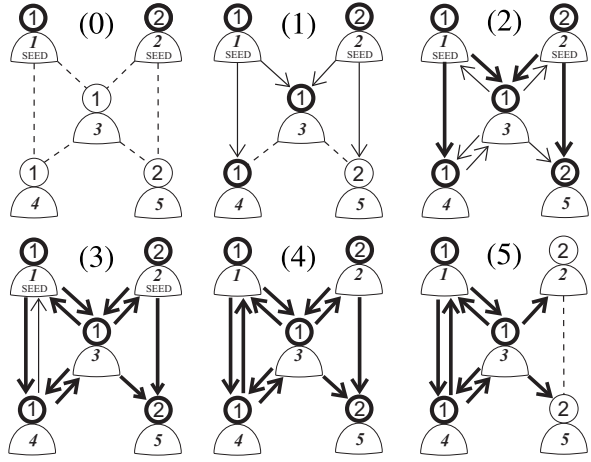}
\end{center}
\caption{Dynamic updates during reversible bootstrap
  percolation. Stage (0): Two seed nodes are activated. The static
  links are shown by dashed lines. Each node has its activation
  threshold on its head, and its identification number on its
  body. Active nodes are shown by bold circles around their
  heads. Seed nodes are active by definition. $Z_0=0$,
  $S_0=2/5$. Stage (1): The influence links from the seeds are
  created, forming set $I_1$. $Z_1=4/12$. Nodes 3 and 4 become active
  because the number of influence links they have received is greater
  than or equal to their thresholds. $S_1=4/5$. Stage (2): Old
  influence links ($I_1$) are shown by bold arrows. New influence
  links ($I_2$) obtained by the rules of the model from $I_1$ are
  shown by thin arrows. Node 3 receives influence from node 4, because
  node 4 does not depend on node 3 to stay at the threshold. However,
  seed 1 does not receive the influence from node 4 because the latter
  does depend on the former to stay at the threshold. In contrast, node
  3 influence all of its neighbors, because, from the point of view of
  any of its neighbors, node 3 is active without its influence. Node 5
  now receives a sufficient number of influence links to become
  activated. $Z_2=9/12$, $S_2=1$. Stage (3): Seed 1 receives influence
  from node 4, because now node 4 has additional influence from node 3
  and stays at the threshold without the influence of seed 1. Node 5
  cannot send influence links to any of its neighbors, because it
  would be inactive without the support of each particular
  neighbor. $Z_3=10/12$, $S_3=1$. Stage (4): The influence network
  comes to a steady state, but both seeds spontaneously lose their
  seed properties. $Z_4=10/12$, $S_4=1$. Stage (5) Node 2 is
  deactivated, because its number of influence links is below the
  threshold. The influence links that come from node 2 are
  removed. Node 1 stays active at the threshold and influences all of
  its neighbors, because it stays at the threshold without the support
  of each of them. Node 5 falls below the threshold and is
  deactivated. The network comes to a new steady state. $Z_5=8/12$,
  $S_5=3/5$.}
\label{schemB}
\end{figure}

In our model of reversible bootstrap percolation, at any moment of
time t, we define a network of directed {\it influence} links, going
along the static links of the network (Fig.~\ref{schemB}). A static
link may correspond to (a) two influence links going in opposite directions, (b) to a single influence
link going in a single direction, or (c) to no influence links at
all. We define $Z_t$ as the number of influence links $N_i(t)$ divided
by the doubled number of static links: $Z_t=N_i(t)/(2N_l)$. The
status of any nonseed node i at any moment of time is defined by the
number of influence links, $k^\prime(t)$, it receives: if this number is
greater or equal than the activation threshold of this node, $k^\prime_i(t)>k^\ast_i$, the node is active; otherwise it is inactive. The seed
nodes are always active; however, like any other nodes, they also
receive influence links, and this number starts to determine their
status as soon as they spontaneously lose their seed property. If, in
contrast a nonseed node $j$ becomes a seed, it immediately activates;
its number of incoming influence links $k^i_j$ does not change, but it
stops determining its status.

 The initial status of the influence network and seed nodes can be
 arbitrary, but for clarity we assume that at $t=0$ they are empty:
 $f(0)=Z(0)=0$. Then some seed nodes spontaneously appear, and the
 dynamic process of activation and self-pruning of the influence
 network begins (Fig~\ref{schemB}). From a social perspective,
 creation of an influence link corresponds to news spreading, while
 its deletion corresponds to fact checking. In principle, the process
 of deletion and creation can be random: at any moment of time, an
 agent represented by a node starts to examine its influence links. It
 selects a static link, determines a neighbor connected by this link,
 and examines the strength of this neighbor belief. If this neighbor
 is a seed, the influence link leading from the seed to this agent is
 created (if it did not exist before). If the neighbor is a nonseed,
 the agent counts the number of influence links, $k^\prime_n(t)$, that
 the neighbor receives excluding the influence link coming from the
 agent. This exclusion is crucial for the fact checking: the agent
 wants to be sure that the neighbor has a firm belief without an
 influence from the agent. This exclusion corresponds to the word
 ``already'' in the definition of $Z$ in
 Ref. \cite{DiMuroBootstrap}. In Eq.~(\ref{e:Phi}) the exclusion is
 reflected by the reduction of the index of summation from $k$ to
 $k-1$ in the inner sum over $j$; here $k-1$ is the number of outgoing
 links of the neighbor, excluding one link by which this neighbor was
 reached from the agent.  If $k^\prime_n \geq k^\ast_n$, the agent
 creates an influence link from this neighbor, if it did not exist
 before. If in contrast, $k^\prime_n <k^\ast_n$, the agent deletes an
 influence link from this neighbor, if it did exist. By this method,
 which represents both news spreading and fact checking, the
 influence network is kept updated. It is obvious that this method
 exactly corresponds to the iterative Eq.~(\ref{iter}), if at each
 time step $n$ the status of all influence links is checked
 simultaneously. Computationally, we keep two arrays of influence
 links $I_{n-1}$ and
 \noindent $I_n$; array $I_n$ at first is empty and is
 created using existing array $I_{n-1}$. Once $I_n$ is created, the
 fraction of influence links $Z_n$ is computed and the set $A_n$ of
 active nodes is determined as the nodes for which the number of
 influence links they receive is $k^i_i\geq k^\ast_i$.  This
 determination corresponds to the computation of the fraction of the
 active nodes $S_n$ using Eq.~(\ref{e:Sn}). Note that in the direct
 bootstrap process, described above, $k^a_i=k^i_i$, and there is no
 need to introduce the influence network to describe the activation
 process of nodes starting from $Z_0=0, S_0=f$. However, once the
 fraction of seed nodes changes, especially if it decreases, the
 network of influence links must be used to construct a reversible
 bootstrap process, in which the fraction of active nodes reduces in
 response to the reduction of the number of seeds. We see that the
 dynamic process of the influence link updates together with the set
 of seeds completely determines the set of active nodes at each
 stage. Conversely, the influence network can be constructed for any
 initial set of active nodes, $A_0$, and a set of seed nodes, $F_0$, by
 the same update process described above if we begin our iterations
 with an influence network consisting of directed links emanating from
 each active node. During this process which resembles pruning,
 certain influence links are removed; as a result, certain nodes are
 deactivated and the influence network converges to a steady state,
 described by Eqs.~(\ref{self}) and (\ref{Act}).

 If the seed nodes change infrequently compared to the influence network update, that is, after each change $Z_n$ stabilizes to one of the stable fixed points of Eq.~(\ref{self}), then at each stage $n$ the values of $Z_n$ and $S_n$ predicted by Eqs. (\ref{iter}) and (\ref{e:Sn}) coincide with the simulation results for a large enough network.

 If the seeds change frequently before $Z_n$ stabilizes, we have a dynamical process similar to one presented in \cite{Majdandzic2014}, with the difference that reversible bootstrap satisfies exact equations for
 $\Phi$ and $\Psi$, while in \cite{Majdandzic2014} the equations are approximate in the limit of large $\langle k\rangle$, because in this limit function $\Psi$, used in Ref.~\cite{Majdandzic2014}, approximates function $\Phi$. The dynamic process introduced here can be applied for many social and natural phenomena and has rich possibilities for generalization such as varying the speed of change of seed nodes, changing the method of updating the influence network, and even a possibility of changing static links and activation thresholds. Also it is obvious that one can construct a reversible $k$-core process using a reversible complementary bootstrap process.

\begin{figure}[H]
\begin{center}
  \includegraphics[width=1\textwidth]{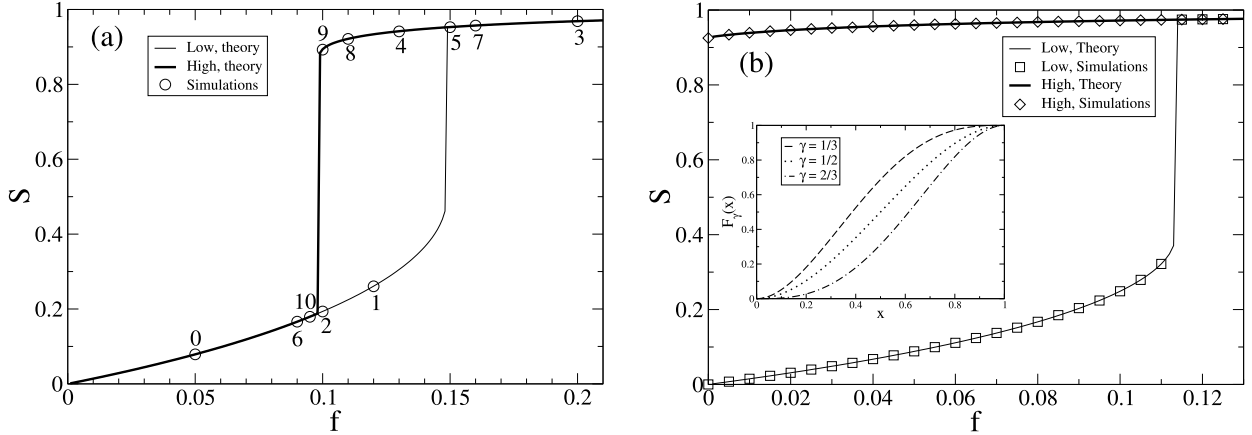}
\end{center}
\caption{Comparison of the theory and simulations for the reversible bootstrap simulation model. In both panels the degree distribution $P(k)$ is Poisson with
  $\langle k\rangle=10$ and $r(j,k)=F_\gamma(j/k)$  with $\gamma=0.53$ (a) and $\gamma=0.48$ (b). In the inset we show the function $F_\gamma(x)$ for $\gamma=1/3$, $\gamma=1/2$, and $\gamma=2/3$. In both cases the size of the network is $N=10^7$ nodes. In (a) the simulations follow a reversible bootstrap process based on the influence network updates described above for a sequence of $f_t$, marked on the graph by numbers $t=0,1,2,...,10$. For (b), the simulations are done for increasing $f$ from $f=0$ to $f=0.125$ giving the low solution and then in reverse order from $f=0.125$ to $f=0$ giving the high solution. One can see that in panel (b) the fraction of active nodes once it undergoes transition to a high activated phase at $f=f_t=0.114$ never returns to a low activated phase even for $f=0$. One can see that a small increase in susceptibility of the population to the fake news, modeled by a small decrease of the inflection point in the distribution $r(j,k)$, may lead to the irreversible opinion change of the population. We obtained similar results for many other types of distributions $P(k)$ and $r(j,k)$.}\label{f:fn}
\end{figure}

\begin{figure}[H]
\begin{center}
  \includegraphics[width=0.8\textwidth]{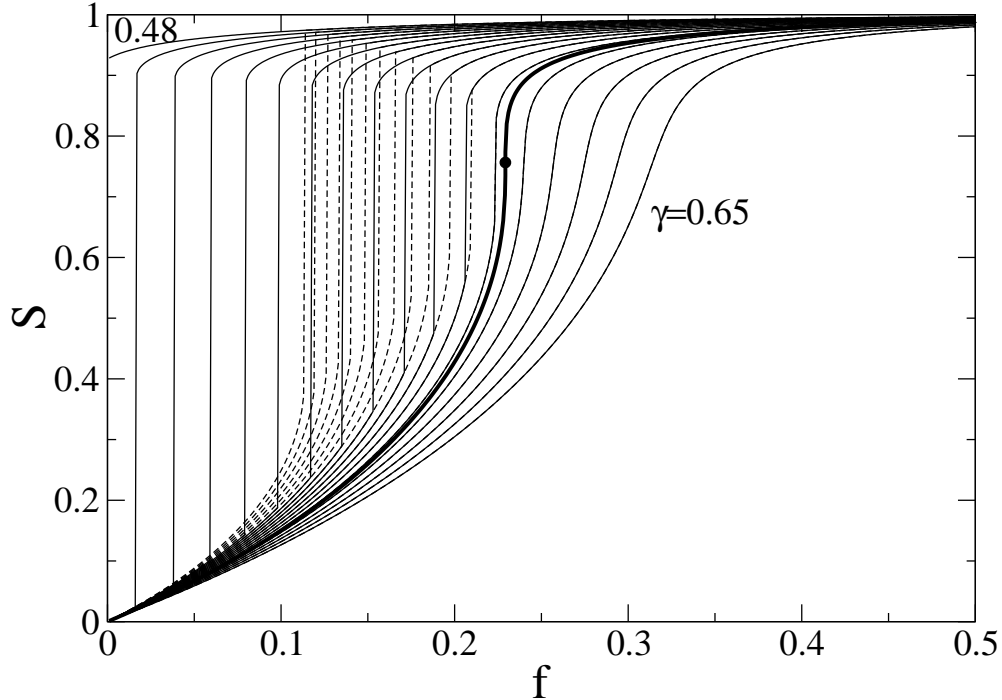}
\end{center}
\caption{Phase diagram of the reversible bootstrap model in the $(f,S)$ plane, which is completely analogous the Van der Waals phase diagram in the $(P,\rho)$ plane. The inflection point
  of the threshold distribution is analogous to temperature. The black line corresponds to isochores (lines of equal $\gamma$ with steps of $.01$ from $\gamma=0.48$ to $\gamma=0.65$) following the high stable point of Eq.~(\ref{iter}), physically analogous to the superheated metastable liquid crossing the liquid-gas spinodal. The dashed lines follow the low stable solution, physically corresponding to the metastable supercooled gas phase crossing the gas-liquid spinodal. Above the critical point $\gamma=0.6035$, $f=0.2292$, $S=0.754$ the second stable fixed point of Eq.~(\ref{iter}) disappears. } 
  \label{diagram}
\end{figure}

The theoretical prediction of the reversible bootstrap percolation
model is in excellent agreement with simulations of finite graphs
(Fig.~\ref{f:fn}). Here we have created a sequence of
\noindent seed-changing
events $f_t$ separated by large enough time intervals so that between
them the network comes to a steady state. That is, once the number of
seeds change it remains the same until the fake news spreading or the
fact checking process finishes. We see that the network strictly
follows the hysteresis loop predicted by the theory. In these
simulations we use a network with the Poisson degree distribution, and
a rather complex threshold distribution $r(j,k)=F_\gamma(j/k)$, where
function
$F_\gamma(x)=[x^2(18\gamma^2-12\gamma)+x^3(4-12\gamma^2)+x^4(6\gamma-3)]/(6\gamma^2-6\gamma+1)$
has an inflection point at $x=\gamma$. The reason for using such a
complex function is twofold. First we want to demonstrate that the
model works for an arbitrary threshold distribution. Indeed, if the
distribution of the activation thresholds can be obtained empirically
for certain social or economic networks, the model can use these
empirical distributions. But for this particular $r(j,k)$, which is a
polynomial of $j/k$, the functions $\Phi$ and $\Psi$ can be computed
analytically (see supplementary information of
Ref. \cite{di2017cascading}). Second, we want to demonstrate that a systematic change of
the inflection point of the threshold distribution dramatically
changes the outcome of the model, which has a nontrivial
phase diagram in the $(f,\gamma,S)$ space (Fig.~\ref{diagram}) corresponding
to the phase diagram of the Van der Waals model (see, e.g.,
\cite{Huangbook}) in the $(P,T,\rho)$ space with a critical point, two spinodals,
and the mean-field critical exponents $\beta=1/2$, $\gamma=1$, $\delta=3$. In this case, the
ordering field, pressure $P$, corresponds to the fraction of seeds,
$f$, which is (using political terms) the intensity of fake news
creation. The thermal field, temperature $T$, corresponds to $\gamma$,
which models the resistance of the population to the fake news. A lower
inflection point in the $r(j,k)$ distribution indicates lower
threshold to the fake news. Finally the density $\rho$ is analogous to
$S$ because both are order parameters. A similar phase diagram appears
if we keep $\gamma$ constant but vary $\langle k\rangle$.

A nontrivial outcome of this model is that once the public opinion is
switched by spreading fake news to a new state undergoing a spinodal
crossing, it may not return to the original state even when all the
fake news sources are discarded. Moreover, for certain distributions
$P(k)$ and $r(j,k)$ [Fig.~\ref{f:fn}(b)], the fraction of active nodes
will not undergo a reverse phase transition and will not return to a
low-fraction state, even if all seeds are eliminated, as demonstrated
also in a schematic Fig.~\ref{schemB} for a very small system. In
Fig.~\ref{f:fn}(b) the degree distribution is the same, but the
bootstrap thresholds are slightly lower than in Fig.~\ref{f:fn}(a). In political language, the fake news stories now are closer to the hearts
of the population than in Fig.~\ref{f:fn}(a), so the population
accepts them more easily and keeps believing them even after all of them
have been firmly discarded and all sources of them are deactivated.

In conclusion, we construct and investigate a model of reversible
bootstrap percolation which can be applied to political and social
science, especially to the problem of fake news spreading and
efficiency of fact checking. This model shows that
misconceptions that have been established in a population may prevail,
even when many of the primary sources that disseminated them 
are debunked.

L.A.B. and S.V.B acknowledge support from NSF Grant
No. PHY-1505000 and DTRA Grants No. HDTRA1-14-1-
0017 and No. HDTRA11910016. S.V.B. acknowledge the
partial support of this research through the Dr. Bernard W.
Gamson Computational Science Center at Yeshiva College.
M.A.D. and L.A.B. thank UNMdP (Grant No. EXA 853/18)
and CONICET (Grant No. PIP 00443/2014) for financial
support


\begin{thebibliography}{22}%
\makeatletter
\providecommand \@ifxundefined [1]{%
 \@ifx{#1\undefined}
}%
\providecommand \@ifnum [1]{%
 \ifnum #1\expandafter \@firstoftwo
 \else \expandafter \@secondoftwo
 \fi
}%
\providecommand \@ifx [1]{%
 \ifx #1\expandafter \@firstoftwo
 \else \expandafter \@secondoftwo
 \fi
}%
\providecommand \natexlab [1]{#1}%
\providecommand \enquote  [1]{``#1''}%
\providecommand \bibnamefont  [1]{#1}%
\providecommand \bibfnamefont [1]{#1}%
\providecommand \citenamefont [1]{#1}%
\providecommand \href@noop [0]{\@secondoftwo}%
\providecommand \href [0]{\begingroup \@sanitize@url \@href}%
\providecommand \@href[1]{\@@startlink{#1}\@@href}%
\providecommand \@@href[1]{\endgroup#1\@@endlink}%
\providecommand \@sanitize@url [0]{\catcode `\\12\catcode `\$12\catcode
  `\&12\catcode `\#12\catcode `\^12\catcode `\_12\catcode `\%12\relax}%
\providecommand \@@startlink[1]{}%
\providecommand \@@endlink[0]{}%
\providecommand \url  [0]{\begingroup\@sanitize@url \@url }%
\providecommand \@url [1]{\endgroup\@href {#1}{\urlprefix }}%
\providecommand \urlprefix  [0]{URL }%
\providecommand \Eprint [0]{\href }%
\providecommand \doibase [0]{http://dx.doi.org/}%
\providecommand \selectlanguage [0]{\@gobble}%
\providecommand \bibinfo  [0]{\@secondoftwo}%
\providecommand \bibfield  [0]{\@secondoftwo}%
\providecommand \translation [1]{[#1]}%
\providecommand \BibitemOpen [0]{}%
\providecommand \bibitemStop [0]{}%
\providecommand \bibitemNoStop [0]{.\EOS\space}%
\providecommand \EOS [0]{\spacefactor3000\relax}%
\providecommand \BibitemShut  [1]{\csname bibitem#1\endcsname}%
\let\auto@bib@innerbib\@empty
\bibitem [{\citenamefont {Watts}(2002)}]{Watts}%
  \BibitemOpen
  \bibfield  {author} {\bibinfo {author} {\bibfnamefont {D.~J.}\ \bibnamefont
  {Watts}},\ }\href@noop {} {\bibfield  {journal} {\bibinfo  {journal} {Proc.
  Natl. Acad. Sci. U.S.A.}\ }\textbf {\bibinfo {volume} {99}},\ \bibinfo
  {pages} {5766} (\bibinfo {year} {2002})}\BibitemShut {NoStop}%
\bibitem [{\citenamefont {Chalupa}\ \emph {et~al.}(1979)\citenamefont
  {Chalupa}, \citenamefont {Leath},\ and\ \citenamefont {Reich}}]{Chalupa}%
  \BibitemOpen
  \bibfield  {author} {\bibinfo {author} {\bibfnamefont {J.}~\bibnamefont
  {Chalupa}}, \bibinfo {author} {\bibfnamefont {P.~L.}\ \bibnamefont {Leath}},
  \ and\ \bibinfo {author} {\bibfnamefont {G.~R.}\ \bibnamefont {Reich}},\
  }\href@noop {} {\bibfield  {journal} {\bibinfo  {journal} {J. Phys. C: Solid
  State Phys.}\ }\textbf {\bibinfo {volume} {12}},\ \bibinfo {pages} {L31}
  (\bibinfo {year} {1979})}\BibitemShut {NoStop}%
\bibitem [{\citenamefont {Janson}\ \emph {et~al.}(2009)\citenamefont {Janson}
  \emph {et~al.}}]{janson2009percolation}%
  \BibitemOpen
  \bibfield  {author} {\bibinfo {author} {\bibfnamefont {S.}~\bibnamefont
  {Janson}} \emph {et~al.},\ }\href@noop {} {\bibfield  {journal} {\bibinfo
  {journal} {Electron. J. Probab.}\ }\textbf {\bibinfo {volume} {14}},\
  \bibinfo {pages} {86} (\bibinfo {year} {2009})}\BibitemShut {NoStop}%
\bibitem [{\citenamefont {Baxter}\ \emph {et~al.}(2010)\citenamefont {Baxter},
  \citenamefont {Dorogovtsev}, \citenamefont {Goltsev},\ and\ \citenamefont
  {Mendes}}]{Bax_2010}%
  \BibitemOpen
  \bibfield  {author} {\bibinfo {author} {\bibfnamefont {G.~J.}\ \bibnamefont
  {Baxter}}, \bibinfo {author} {\bibfnamefont {S.~N.}\ \bibnamefont
  {Dorogovtsev}}, \bibinfo {author} {\bibfnamefont {A.~V.}\ \bibnamefont
  {Goltsev}}, \ and\ \bibinfo {author} {\bibfnamefont {J.~F.~F.}\ \bibnamefont
  {Mendes}},\ }\href@noop {} {\bibfield  {journal} {\bibinfo  {journal} {Phys.
  Rev. E}\ }\textbf {\bibinfo {volume} {82}},\ \bibinfo {pages} {011103}
  (\bibinfo {year} {2010})}\BibitemShut {NoStop}%
\bibitem [{\citenamefont {Baxter}\ \emph {et~al.}(2011)\citenamefont {Baxter},
  \citenamefont {Dorogovtsev}, \citenamefont {Goltsev},\ and\ \citenamefont
  {Mendes}}]{Bax_2011}%
  \BibitemOpen
  \bibfield  {author} {\bibinfo {author} {\bibfnamefont {G.~J.}\ \bibnamefont
  {Baxter}}, \bibinfo {author} {\bibfnamefont {S.~N.}\ \bibnamefont
  {Dorogovtsev}}, \bibinfo {author} {\bibfnamefont {A.~V.}\ \bibnamefont
  {Goltsev}}, \ and\ \bibinfo {author} {\bibfnamefont {J.~F.~F.}\ \bibnamefont
  {Mendes}},\ }\href@noop {} {\bibfield  {journal} {\bibinfo  {journal} {Phys.
  Rev. E}\ }\textbf {\bibinfo {volume} {83}},\ \bibinfo {pages} {051134}
  (\bibinfo {year} {2011})}\BibitemShut {NoStop}%
\bibitem [{\citenamefont {Centola}(2010)}]{centola2010spread}%
  \BibitemOpen
  \bibfield  {author} {\bibinfo {author} {\bibfnamefont {D.}~\bibnamefont
  {Centola}},\ }\href@noop {} {\bibfield  {journal} {\bibinfo  {journal}
  {Science}\ }\textbf {\bibinfo {volume} {329}},\ \bibinfo {pages} {1194}
  (\bibinfo {year} {2010})}\BibitemShut {NoStop}%
\bibitem [{\citenamefont {Jackson}\ and\ \citenamefont
  {{L\'opez-Pintado}}(2013)}]{JacksonDiff}%
  \BibitemOpen
  \bibfield  {author} {\bibinfo {author} {\bibfnamefont {M.~O.}\ \bibnamefont
  {Jackson}}\ and\ \bibinfo {author} {\bibfnamefont {D.}~\bibnamefont
  {{L\'opez-Pintado}}},\ }\href@noop {} {\bibfield  {journal} {\bibinfo
  {journal} {Netw. Sci.}\ }\textbf {\bibinfo {volume} {1}},\ \bibinfo {pages}
  {49} (\bibinfo {year} {2013})}\BibitemShut {NoStop}%
\bibitem [{\citenamefont {Kempe}\ \emph {et~al.}(2003)\citenamefont {Kempe},
  \citenamefont {Kleinberg},\ and\ \citenamefont
  {Tardos}}]{kempe2003maximizing}%
  \BibitemOpen
  \bibfield  {author} {\bibinfo {author} {\bibfnamefont {D.}~\bibnamefont
  {Kempe}}, \bibinfo {author} {\bibfnamefont {J.}~\bibnamefont {Kleinberg}}, \
  and\ \bibinfo {author} {\bibfnamefont {{\'E}.}~\bibnamefont {Tardos}},\ }in\
  \href@noop {} {\emph {\bibinfo {booktitle} {Proceedings of the Ninth {ACM
  SIGKDD} International Conference on Knowledge Discovery and Data Mining}}}\
  (\bibinfo {organization} {ACM},\ \bibinfo {address} {New York},\ \bibinfo
  {year} {August 2003})\ pp.\ \bibinfo {pages} {137--146}\BibitemShut {NoStop}%
\bibitem [{\citenamefont {Domingos}\ and\ \citenamefont
  {Richardson}(2001)}]{DomingosMarketing}%
  \BibitemOpen
  \bibfield  {author} {\bibinfo {author} {\bibfnamefont {P.}~\bibnamefont
  {Domingos}}\ and\ \bibinfo {author} {\bibfnamefont {M.}~\bibnamefont
  {Richardson}},\ }in\ \href@noop {} {\emph {\bibinfo {booktitle} {Proceedings
  of the Seventh {ACM SIGKDD} International Conference on Knowledge Discovery
  and Data Mining}}}\ (\bibinfo {organization} {ACM},\ \bibinfo {year} {August
  2001})\ pp.\ \bibinfo {pages} {57--66}\BibitemShut {NoStop}%
\bibitem [{\citenamefont {Shrestha}\ and\ \citenamefont
  {Moore}(2014)}]{Shrestha}%
  \BibitemOpen
  \bibfield  {author} {\bibinfo {author} {\bibfnamefont {M.}~\bibnamefont
  {Shrestha}}\ and\ \bibinfo {author} {\bibfnamefont {C.}~\bibnamefont
  {Moore}},\ }\href@noop {} {\bibfield  {journal} {\bibinfo  {journal} {Phys.
  Rev. E}\ }\textbf {\bibinfo {volume} {89}},\ \bibinfo {pages} {022805}
  (\bibinfo {year} {2014})}\BibitemShut {NoStop}%
\bibitem [{\citenamefont {Gleeson}(2008)}]{Gleesoncorrelated}%
  \BibitemOpen
  \bibfield  {author} {\bibinfo {author} {\bibfnamefont {J.~P.}\ \bibnamefont
  {Gleeson}},\ }\href@noop {} {\bibfield  {journal} {\bibinfo  {journal} {Phys.
  Rev. E}\ }\textbf {\bibinfo {volume} {77}},\ \bibinfo {pages} {046117}
  (\bibinfo {year} {2008})}\BibitemShut {NoStop}%
\bibitem [{\citenamefont {Kempe}\ \emph {et~al.}(2005)\citenamefont {Kempe},
  \citenamefont {Kleinberg},\ and\ \citenamefont {Tardos}}]{Kempe_Inf}%
  \BibitemOpen
  \bibfield  {author} {\bibinfo {author} {\bibfnamefont {D.}~\bibnamefont
  {Kempe}}, \bibinfo {author} {\bibfnamefont {J.}~\bibnamefont {Kleinberg}}, \
  and\ \bibinfo {author} {\bibfnamefont {{\'E}.}~\bibnamefont {Tardos}},\ }in\
  \href@noop {} {\emph {\bibinfo {booktitle} {Automata, Languages and
  Programming}}}\ (\bibinfo  {publisher} {Springer Berlin},\ \bibinfo {year}
  {2005})\ pp.\ \bibinfo {pages} {1127--1138}\BibitemShut {NoStop}%
\bibitem [{\citenamefont {Kirkpatrick}\ \emph {et~al.}(2002)\citenamefont
  {Kirkpatrick}, \citenamefont {Winfried}, \citenamefont {Robert},\ and\
  \citenamefont {Harald}}]{Kirkpatrick}%
  \BibitemOpen
  \bibfield  {author} {\bibinfo {author} {\bibfnamefont {S.}~\bibnamefont
  {Kirkpatrick}}, \bibinfo {author} {\bibfnamefont {W.~W.}\ \bibnamefont
  {Winfried}}, \bibinfo {author} {\bibfnamefont {G.~B.}\ \bibnamefont
  {Robert}}, \ and\ \bibinfo {author} {\bibfnamefont {H.}~\bibnamefont
  {Harald}},\ }\href@noop {} {\bibfield  {journal} {\bibinfo  {journal}
  {Physica A}\ }\textbf {\bibinfo {volume} {314}} (\bibinfo {year}
  {2002})}\BibitemShut {NoStop}%
\bibitem [{\citenamefont {Granell}\ \emph {et~al.}(2014)\citenamefont
  {Granell}, \citenamefont {G\'omez},\ and\ \citenamefont {Arenas}}]{Granell}%
  \BibitemOpen
  \bibfield  {author} {\bibinfo {author} {\bibfnamefont {C.}~\bibnamefont
  {Granell}}, \bibinfo {author} {\bibfnamefont {S.}~\bibnamefont {G\'omez}}, \
  and\ \bibinfo {author} {\bibfnamefont {A.}~\bibnamefont {Arenas}},\
  }\href@noop {} {\bibfield  {journal} {\bibinfo  {journal} {Phys. Rev. E}\
  }\textbf {\bibinfo {volume} {90}},\ \bibinfo {pages} {012808} (\bibinfo
  {year} {2014})}\BibitemShut {NoStop}%
\bibitem [{\citenamefont {Feng}\ \emph {et~al.}(2015)\citenamefont {Feng},
  \citenamefont {Hu}, \citenamefont {Li}, \citenamefont {Stanley},
  \citenamefont {Havlin},\ and\ \citenamefont {Braunstein}}]{Feng}%
  \BibitemOpen
  \bibfield  {author} {\bibinfo {author} {\bibfnamefont {L.}~\bibnamefont
  {Feng}}, \bibinfo {author} {\bibfnamefont {Y.}~\bibnamefont {Hu}}, \bibinfo
  {author} {\bibfnamefont {B.}~\bibnamefont {Li}}, \bibinfo {author}
  {\bibfnamefont {H.~E.}\ \bibnamefont {Stanley}}, \bibinfo {author}
  {\bibfnamefont {S.}~\bibnamefont {Havlin}}, \ and\ \bibinfo {author}
  {\bibfnamefont {L.~A.}\ \bibnamefont {Braunstein}},\ }\href@noop {}
  {\bibfield  {journal} {\bibinfo  {journal} {PLOS ONE}\ }\textbf {\bibinfo
  {volume} {10}},\ \bibinfo {pages} {1} (\bibinfo {year} {2015})}\BibitemShut
  {NoStop}%
\bibitem [{\citenamefont {Huang}(2009)}]{Huangbook}%
  \BibitemOpen
  \bibfield  {author} {\bibinfo {author} {\bibfnamefont {K.}~\bibnamefont
  {Huang}},\ }\href@noop {} {\emph {\bibinfo {title} {{\it Introduction to
  Statistical Physics}}}},\ \bibinfo {edition} {2nd}\ ed.\ (\bibinfo
  {publisher} {{CRC} Press, Boca Raton},\ \bibinfo {year} {2009})\BibitemShut
  {NoStop}%
\bibitem [{\citenamefont {{Di Muro}}\ \emph {et~al.}(2019)\citenamefont {{Di
  Muro}}, \citenamefont {Valdez}, \citenamefont {Stanley}, \citenamefont
  {Buldyrev},\ and\ \citenamefont {Braunstein}}]{DiMuroBootstrap}%
  \BibitemOpen
  \bibfield  {author} {\bibinfo {author} {\bibfnamefont {M.~A.}\ \bibnamefont
  {{Di Muro}}}, \bibinfo {author} {\bibfnamefont {L.~D.}\ \bibnamefont
  {Valdez}}, \bibinfo {author} {\bibfnamefont {H.~E.}\ \bibnamefont {Stanley}},
  \bibinfo {author} {\bibfnamefont {S.~V.}\ \bibnamefont {Buldyrev}}, \ and\
  \bibinfo {author} {\bibfnamefont {L.~A.}\ \bibnamefont {Braunstein}},\
  }\href@noop {} {\bibfield  {journal} {\bibinfo  {journal} {Phys. Rev. E}\
  }\textbf {\bibinfo {volume} {99}},\ \bibinfo {pages} {022311} (\bibinfo
  {year} {2019})}\BibitemShut {NoStop}%
\bibitem [{\citenamefont {Min}\ and\ \citenamefont {Goh}(2014)}]{Min2014}%
  \BibitemOpen
  \bibfield  {author} {\bibinfo {author} {\bibfnamefont {B.}~\bibnamefont
  {Min}}\ and\ \bibinfo {author} {\bibfnamefont {K.~I.}\ \bibnamefont {Goh}},\
  }\href@noop {} {\bibfield  {journal} {\bibinfo  {journal} {Phys. Rev. E}\
  }\textbf {\bibinfo {volume} {89}},\ \bibinfo {pages} {040802(R)} (\bibinfo
  {year} {2014})}\BibitemShut {NoStop}%
\bibitem [{\citenamefont {Majdandzic}\ \emph {et~al.}(2014)\citenamefont
  {Majdandzic}, \citenamefont {Podobnik}, \citenamefont {Buldyrev},
  \citenamefont {Kenett}, \citenamefont {Havlin},\ and\ \citenamefont
  {Stanley}}]{Majdandzic2014}%
  \BibitemOpen
  \bibfield  {author} {\bibinfo {author} {\bibfnamefont {A.}~\bibnamefont
  {Majdandzic}}, \bibinfo {author} {\bibfnamefont {B.}~\bibnamefont
  {Podobnik}}, \bibinfo {author} {\bibfnamefont {S.~V.}\ \bibnamefont
  {Buldyrev}}, \bibinfo {author} {\bibfnamefont {D.~Y.}\ \bibnamefont
  {Kenett}}, \bibinfo {author} {\bibfnamefont {S.}~\bibnamefont {Havlin}}, \
  and\ \bibinfo {author} {\bibfnamefont {H.~E.}\ \bibnamefont {Stanley}},\
  }\href@noop {} {\bibfield  {journal} {\bibinfo  {journal} {Nat. Phys.}\
  }\textbf {\bibinfo {volume} {10}},\ \bibinfo {pages} {34} (\bibinfo {year}
  {2014})}\BibitemShut {NoStop}%
\bibitem [{\citenamefont {Dorogovtsev}\ \emph {et~al.}(2006)\citenamefont
  {Dorogovtsev}, \citenamefont {Goltsev},\ and\ \citenamefont
  {Mendes}}]{dorogovtsev2006k}%
  \BibitemOpen
  \bibfield  {author} {\bibinfo {author} {\bibfnamefont {S.~N.}\ \bibnamefont
  {Dorogovtsev}}, \bibinfo {author} {\bibfnamefont {A.~V.}\ \bibnamefont
  {Goltsev}}, \ and\ \bibinfo {author} {\bibfnamefont {J.~F.~F.}\ \bibnamefont
  {Mendes}},\ }\href@noop {} {\bibfield  {journal} {\bibinfo  {journal} {Phys.
  Rev. Lett.}\ }\textbf {\bibinfo {volume} {96}},\ \bibinfo {pages} {040601}
  (\bibinfo {year} {2006})}\BibitemShut {NoStop}%
\bibitem [{\citenamefont {{Azimi-Tafreshi}}\ \emph {et~al.}(2019)\citenamefont
  {{Azimi-Tafreshi}}, \citenamefont {Osat},\ and\ \citenamefont
  {Dorogovtsev}}]{doro_2019}%
  \BibitemOpen
  \bibfield  {author} {\bibinfo {author} {\bibfnamefont {N.}~\bibnamefont
  {{Azimi-Tafreshi}}}, \bibinfo {author} {\bibfnamefont {S.}~\bibnamefont
  {Osat}}, \ and\ \bibinfo {author} {\bibfnamefont {S.~N.}\ \bibnamefont
  {Dorogovtsev}},\ }\href@noop {} {\bibfield  {journal} {\bibinfo  {journal}
  {Phys. Rev. E}\ }\textbf {\bibinfo {volume} {99}},\ \bibinfo {pages} {022312}
  (\bibinfo {year} {2019})}\BibitemShut {NoStop}%
\bibitem [{\citenamefont {{Di Muro}}\ \emph {et~al.}(2017)\citenamefont {{Di
  Muro}}, \citenamefont {Valdez}, \citenamefont {{Arag{\~a}o R{\^e}go}},
  \citenamefont {Buldyrev}, \citenamefont {Stanley},\ and\ \citenamefont
  {Braunstein}}]{di2017cascading}%
  \BibitemOpen
  \bibfield  {author} {\bibinfo {author} {\bibfnamefont {M.~A.}\ \bibnamefont
  {{Di Muro}}}, \bibinfo {author} {\bibfnamefont {L.~D.}\ \bibnamefont
  {Valdez}}, \bibinfo {author} {\bibfnamefont {H.~H.}\ \bibnamefont
  {{Arag{\~a}o R{\^e}go}}}, \bibinfo {author} {\bibfnamefont {S.~V.}\
  \bibnamefont {Buldyrev}}, \bibinfo {author} {\bibfnamefont {H.~E.}\
  \bibnamefont {Stanley}}, \ and\ \bibinfo {author} {\bibfnamefont {L.~A.}\
  \bibnamefont {Braunstein}},\ }\href@noop {} {\bibfield  {journal} {\bibinfo
  {journal} {Sci. Rep.}\ }\textbf {\bibinfo {volume} {7}},\ \bibinfo {pages}
  {15059} (\bibinfo {year} {2017})}\BibitemShut {NoStop}%
\end{thebibliography}

%

\end{document}